\documentclass[a4paper,aps,english,twocolumn,floatfix,amsfonts,amssymb,superscriptaddress]{revtex4}
\usepackage{epstopdf}
\usepackage{graphicx}
\usepackage{dcolumn}
\usepackage{bm}
\usepackage{bbm,bbold}
\usepackage{color}
\usepackage{xcolor, soul}
\sethlcolor{green}
\usepackage{amsfonts}
\usepackage{amsmath}
\usepackage{mdframed}
\usepackage[normalem]{ulem}
\usepackage{mathrsfs}
\usepackage[percent]{overpic}
\usepackage[colorlinks=true,citecolor=blue]{hyperref}
\hypersetup{colorlinks=true,citecolor=blue,linkcolor=blue,urlcolor=blue}
\let\oldAA\AA
\renewcommand{\AA}{\text{\normalfont\oldAA}}

\begin{document}
\title{Exciton routing in the heterostructure of TMD and paraelectrics}

\author{V. Shahnazaryan}
\affiliation{ITMO University, St. Petersburg 197101, Russia}

\author{O. Kyriienko}
\affiliation{NORDITA, KTH Royal Institute of Technology and Stockholm University,
Roslagstullsbacken 23, SE-106 91 Stockholm, Sweden}
\affiliation{ITMO University, St. Petersburg 197101, Russia}
\author{H. Rostami}
\affiliation{NORDITA, KTH Royal Institute of Technology and Stockholm University,
Roslagstullsbacken 23, SE-106 91 Stockholm, Sweden}
\begin{abstract}
We propose a scheme for the spatial exciton energy control and exciton routing in a transition metal dichalcogenide (TMD) monolayer which lies on a quantum paraelectric substrate. It relies on the ultrasensitive response of the substrate dielectric permittivity to temperature changes, allowing for spatially inhomogeneous screening of Coulomb interaction in a monolayer. As an example, we consider the heterostructure of TMD and strontium titanate oxide SrTiO$_3$, where large dielectric screening can be attained. We study the impact of substrate temperature on the characteristic electronic features of TMD monolayers such as the particle bandgap  and   exciton binding energy, Bohr radius and nonlinearity (an exciton-exciton interaction). The combination of particle bandgap and exciton binding energy modulation results in the shift of the exciton resonance energy.
Applying local heating, we create spatial patterns with varying exciton resonant energy and an exciton flow towards the energetically lower region of the sample.
\end{abstract}
\maketitle
\section{Introduction}
The synthesis of atomically thin transition metal dichalcogenides (TMDs), such as MoS$_2$ and WS$_2$, opened new horizons for contemporary semiconductor optics \cite{Mak2010,Novoselov2016}. Thanks to direct band gap and rich diversity of exciton states, caused by the presence of a valley degree of freedom,
the family of TMD monolayers becomes especially favorable for addressing a wide range of exciton-related phenomena \cite{ChernikovReview}. In addition, the TMD monolayers possess giant exciton binding energy of the order of 0.3-0.5 eV, allowing for a room temperature operation. The study of excitonic effects in TMD monolayers covers the spectroscopic measurement \cite{Zhao2013,Shan2014,Zhu2015,Shang2015,Hill2015,Ceballos2016,Schmidt2016,Chernikov2014} and first principle calculation of binding energy \cite{Komsa2012,Rama2012,Lambrecht2012,Yakobson2013,Berkelbach2013,Qiu2013,Komsa2015,Qiu2016}, together with analytic calculations of excitonic properties \cite{Berghauser2014,Glazov2014,Wu2015,Li2015}.

The electronic properties of all two-dimensional (2D) materials, with TMDs being a prominent example, are extremely sensitive to external probes such as external gating potential, local strain, and substrate effect. This circumstance opens a way to the control of exciton resonant energy, arising via the interplay of modification of both electronic bandgap ($E_g$) and exciton binding energy ($E_b$).
In particular, the screening of inter-particle interactions via electrically injected free charge carriers \cite{Chernikov2015} or by the choice of a suitable substrate \cite{Ugeda2014,Lin2014,Raja2017,Borghardt2017,Hsu2019} can be employed in order to manipulate the exciton energy ($E_{\rm X}=E_g-E_b$). It should be mentioned, that the modification of electronic bandgap ($\delta E_g$) and exciton binding energy ($\delta E_b$) have the same sign and therefore the overall change in the exciton energy, $\delta E_{\rm X}=\delta E_g-\delta E_b$, is mostly compensated. Yet, this cancellation is not perfect, resulting in a significant shift of the exciton resonant energy position of the order of tens of meVs.

A more challenging task is to attain spatially resolved control over exciton energy. One way to reach this was suggested via application of  spatially inhomogeneous strain to the monolayer, leading to local modification of the bandgap, and thus creating the energy gradients for quasiparticles throughout the sample \cite{Feng_nph_2012,Gomez_nl_2013,Rostami2015,Pablo_prx_2016}. The latter, known as  exciton funnel effect, can be favorable for enhancing the efficiency of solar cells.

In this paper, we propose a new method in order to maintain spatial control over exciton resonant energy in TMD monolayers. The method is based on a spatially resolved modulation of Coulomb interaction by means of dielectric environment. For this purpose, we employ quantum paraelectric materials such as SrTiO$_3$ (STO) and KTaO$_3$ (KTO) known for  outstanding dielectric properties, as the substrate.

There is a growing interest in exploring the effect of substrate on the electronic and optical properties of 2D materials.  Particularly, the heterostructure of STO substrate with graphene~\cite{Couto_prl_2011,Saha_srep_2014,Ryu_nl_2017,Sahoo_npj_2018,Chen_pssrrl_2018} and TMD~\cite{Zhang_2016,Chen_2018}  is attracting considerable attention. The large dielectric constant of STO substrate strongly screens the long-range Coulomb interactions inside 2D materials. This property has been already employed in experimental transport studies of graphene/STO heterostructure \cite{Couto_prl_2011,Chen_pssrrl_2018}. The angle-resolved spectroscopy of  graphene on STO demonstrated the temperature-dependent nonlinearity of the energy spectrum, which can be attributed to the modulation of the electron-electron interaction as a function of temperature \cite{Ryu_nl_2017}. Recently, it was experimentally shown that  single-layer MoS$_2$ can be grown on STO substrate with different interface terminations \cite{Chen_2018}.
The experimental evidence implies that large triangle-shape flakes MoS$_2$, with up to 10 $\mu$m side lengths, are attached to STO surface owing to a weak van der Waals interaction \cite{Chen_2018}. The experimental measurements show that the photoluminescence efficiency of single-layer MoS$_2$/STO is high enough and this heterostructure could be promising for optoelectronics.

Here, the substrate induced screening is utilized to manipulate the electronic bandgap and characteristic properties of excitons in TMD, {\it e.g.} exciton binding energy and Bohr radius. The latter ultimately results in the modulation of exciton lifetime and nonlinearity, caused by exciton-exciton Coulomb interaction. Precisely, we aim to study the effect of spatially inhomogeneous temperature profiles in STO substrate on the exciton properties in TMD. We show that the spatially resolved heating of the substrate results in a corresponding gradient of the exciton energy. The latter plays a role of the drifting force, {\it routing} the excitons towards the minimum of exciton resonance energy $E_X$, whereas typically the exciton transport in TMD monolayers is of diffusive character \cite{Kumar2014,Mouri2014,Yuan2017,Kulig2018,Cadiz2018,Glazov2019}.

\begin{figure}[t!]
    \centering
    \includegraphics[width=1.\linewidth]{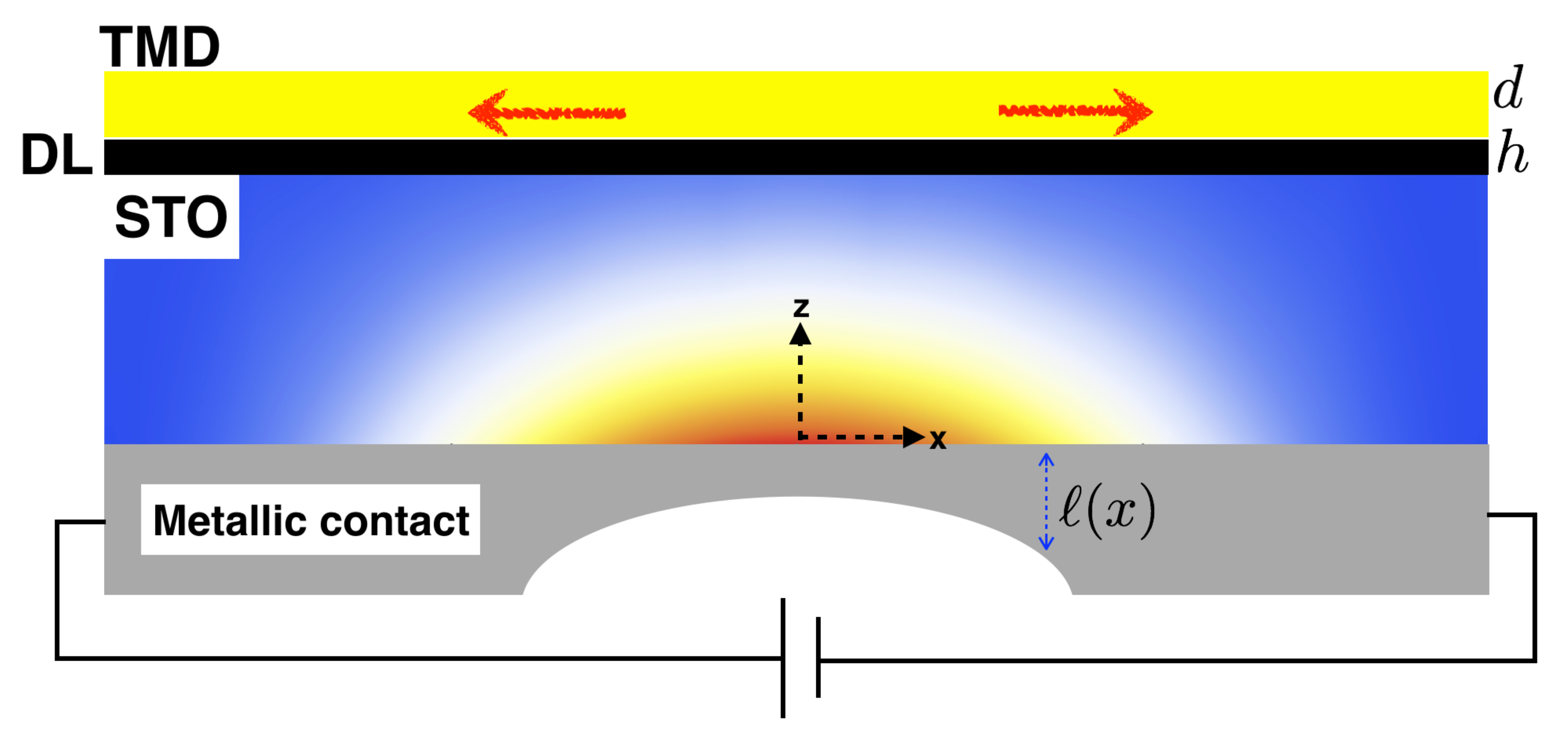}
    \caption{ The sketch of monolayer TMD deposited on top of STO substrate. The substrate is inhomogeneously heated from the bottom by Ohmic contact of varying thickness. The color indicates the variation of temperature along STO substrate.
    In the highly heated region, the substrate permittivity is lower, resulting in strong bandgap renormalization and higher value of exciton resonance energy. The spatially varying landscape of resonance energy routes excitons excited by optical pumping to the lower temperature regions of the sample, allowing to create a controllable current of exciton cloud. The red arrows denotes direction of the exciton gas flow.
    }
    \label{fig:sketch}
\end{figure}

The characteristic feature of the quantum paraelectric compounds ({\it e.g.} STO) is the strong quantum fluctuations close to the ferroelectric critical point which suppresses the ferroelectric order. Inverse dielectric function of these materials scales as $1/\varepsilon(T)\sim T^2$ close to the ferroelectric critical point~\cite{Rowley_np_2014,Fujishita_jpsj_2016}. Due to the quantum paraelectric nature of STO at low temperature, its dielectric constant can reach very high values ($\sim 10^4$) in a few hundred Kelvin change of temperature. Phenomenologically, one can use the Barrett formula for STO's dielectric constant above the critical temperature~\cite{Sawaguchi1962}
\begin{equation}\label{eq:sto}
    \varepsilon_{\rm STO}(T)=\frac{a}{\coth(T_\circ/T)-b},
\end{equation}
where $a\approx 2143$, $b\approx 0.90$, $T_\circ \approx 42$ K. The strong temperature-dependent dielectric constant of STO substrate has a dramatic impact on the Coulomb interaction between charged particles inside the 2D material.
The temperature gradient in STO can be generated in several ways such as local heating by using lasers and Joule heating~\cite{Watelot_prl_2012}. Particularly, in Ref.~\cite{Watelot_prl_2012}, it is shown that by  utilizing Joule heating in a nanocontact, it is possible to create large temperature gradient, $\Delta T\sim 60$ K, within nanoscale depth ($\sim250$ nm). Here, we assume a similar scenario where a desired temperature profile can be achieved in STO by properly designing the thickness of a metallic contact and fabricating a nanocontact. A non-uniform resistivity, ${\cal R}(x)\propto 1/\ell(x)$, comes as a result of the non-uniform thickness of the metallic nanocontact, $\ell(x)$. Considering a current flow in the Ohmic nanocontact with an inhomogeneous resistivity, an inhomogeneous temperature can be achieved at the interface of STO and the metallic contact. A schematic illustration of a non-uniform temperature distribution is shown in Fig.~\ref{fig:sketch}.

The rest of the paper is organized in three sections. In Sec.~\ref{sec:method} we explain the main body of our modeling. In Sec.~\ref{sec:result_discussion} we present our numerical results and corresponding discussion.  In Sec.~\ref{sec:conclusion} we summarize the main achievements of the study, and provide the outlook for future directions.
\section{Methods}\label{sec:method}
We consider a structure of an atomically thin monolayer of MoS$_2$ deposited on a bulk STO substrate, being  a paraelectric material with a giant dielectric constant [see Fig. \ref{fig:sketch}(a)]. The TMD monolayer represents a direct gap semiconductor, which can host excitons at room temperatures.
The system can be described as a two-body problem, which satisfies the following
Schrodinger equation:
\begin{align}
    \label{eq:SchroTot}
    &\left[ -\frac{\hbar^2 \nabla^2_{\bm R}}{2M}  -\frac{\hbar^2 \nabla^2_{\bm r}}{2\mu} +E_g(T({\bm R})) + V({\bm r},T({\bm R}))  \right]  \Psi ({\bm r},{\bm R})
   \nonumber\\
   &
    = E_{\rm X}({\bm R}) \Psi ({\bm r},{\bm R}),
\end{align}
where ${\bm r}$, ${\bm R}$ are the electron-hole relative and center of mass (CM) coordinates,  and $M=m_e+m_h$, $\mu=m_e m_h/M$ are exciton total and reduced masses, respectively. For monolayer MoS$_2$ the electron and hole effective masses are $m_e=0.35m_0$ and $m_h=0.45m_0$, where $m_0$ is the free electron mass \cite{Lambrecht2012}. Here $E_g(T({\bm R}))$ stands for the single-layer bandgap. Evidently, both the bandgap and the exciton binding energy are determined by the Coulomb interaction in the structure. Hence, we first proceed with the analysis of the Coulomb interaction in the system.

Due to the discontinuity of dielectric permittivity at the interfaces of monolayer with substrate and cover layer (or vacuum), the Coulomb interaction $V({\bm r},T({\bm R}))$ substantially differs from the conventional $1/r$ dependence. Typically it is described by the Keldysh-Rytova formula \cite{Keldysh,Cudazzo2011}.
However, the material properties of STO impose additional peculiarity to the inter-particle interactions in the considered system. Particularly, it is known that the capacitance of nanoscale STO based capacitors is strongly affected due to the formation of the so-called \emph{dead layer} at the STO-metal interfaces \cite{Spaldin2006,Chang2009}.  The latter is characterized by much smaller dielectric permittivity, caused by the rearrangement of atoms to compensate the strains on the surface of STO. It is still disputable whether the presence of this layer is inherent property of STO or stems from fabrication imperfections. However, accounting for a dead layer impact allows to provide a more realistic description of such interfaces. In recent experiments with graphene grown on STO substrate it was shown that the STO induced screening is severely quenched \cite{Ryu_nl_2017}, which can be attributed to the dielectric constant reduction due to the dead layer. Based on the aforesaid, we model this extra quenching by considering a thin dead layer at the interface of STO/TMD as sketched in Fig.~\ref{fig:sketch}(a). The generic interaction potential in Fourier space is given by
\begin{equation}\label{eq:CPot}
V(q,T)= \frac{v^{\rm eh}_q }{\varepsilon_{\rm eff}(q,T)},
\end{equation}
where the bare {\it attractive} Coulomb interaction between an electron and a hole reads $v^{\rm eh}_q =-2\pi e^2/q$ with $e$ as the elementary charge. The effective substrate-induced nonlocal dielectric function reads \cite{Florian2018}
\begin{align}\label{eq:epsilon_eff}
&\varepsilon_{\rm eff} (q,T)  =
\frac{1+\tanh^2(qd/2)}
{2 \{1 +\frac{\varepsilon_{\mathrm{TOP}}}{\varepsilon_{\mathrm{TMD}}}\tanh(qd/2) \}}
\times\\ &
\frac{
 f(\varepsilon_1,\varepsilon_2,hq)+
\frac{\varepsilon_{\mathrm{DL}}}{\varepsilon_{\mathrm{TMD}}}\tanh (d q) f(\varepsilon_3,\varepsilon_4,hq)
 }{
 f(\varepsilon_{\mathrm{DL}},\varepsilon_{\mathrm{STO}},hq)+
 \frac{\varepsilon_{\mathrm{DL}}}{\varepsilon_{\mathrm{TMD}}} \tanh({d q}/{2})
f(\varepsilon_{\mathrm{STO}},\varepsilon_{\mathrm{DL}},hq)
}\nonumber
\end{align}
where $f(\varepsilon,\varepsilon',x) = \varepsilon + \varepsilon'\tanh(x)$ and
\begin{align}
&\varepsilon_1=\varepsilon_{\mathrm{DL}}(\varepsilon_{\mathrm{STO}}+ \varepsilon_{\mathrm{TOP}})
~~,~~
\varepsilon_2= \varepsilon_{\mathrm{STO}} \varepsilon_{\mathrm{TOP}}+\varepsilon_{\mathrm{DL}}^2~,
\\&
\varepsilon_3= \varepsilon_{\mathrm{STO}} \varepsilon_{\mathrm{TOP}}+\varepsilon_{\mathrm{TMD}}^2~~,~~
\varepsilon_4= \frac{\varepsilon_{\mathrm{STO}} \varepsilon^2_{\mathrm{TMD}}}{\varepsilon_{\mathrm{DL}}} +\varepsilon_{\mathrm{DL}} \varepsilon_{\mathrm{TOP}}~.
\nonumber
\end{align}
For the shorthand notation in the above formula, we use  $\varepsilon_{\rm STO}$ instead of $\varepsilon_{\rm STO}(T)$.  Notice that $\varepsilon_{\rm TOP} \approx 1$ corresponds to the dielectric constant of air;  $\varepsilon_{\rm TMD}$ and $d$ denote the dielectric constant and the thickness of single-layer TMD. $\varepsilon_{\mathrm{DL}}$ and $h$ correspond to the dielectric permittivity and thickness of dead layer. While the exact values of parameters for the dead layer depend on particular experimental realization, we take its width be of the order of 4$\rm\AA$. The dielectric permittivity of the dead layer is shown to be temperature independent \cite{Chang2009} and much smaller than that of bulk STO, and we assume it to vary in the range of 10 to 100. Here we also disregard the vertical variation of STO temperature. The latter is explained by the evidence that only a few nm thick upper skin of substrate affects  the strength of Coulomb interaction inside the single layer TMD, and the temperature variation on that length scale is negligible.

\begin{figure}[t]
    \centering
    \includegraphics[width=1.\linewidth]{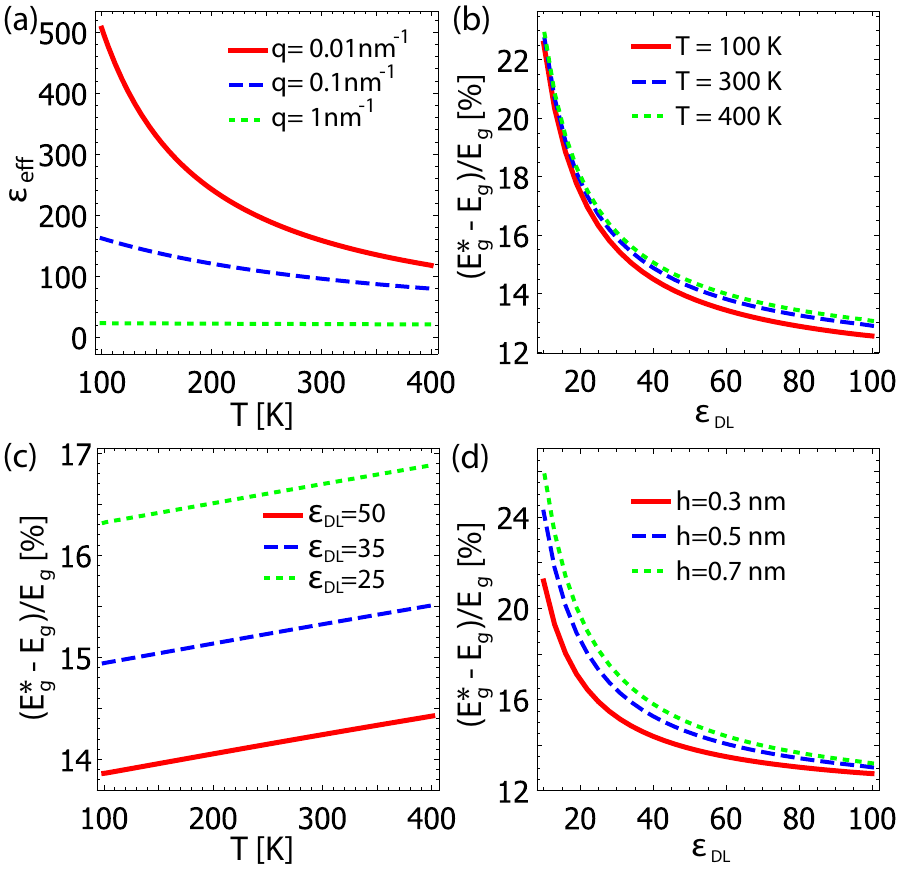}
    \caption{(a) Effective nonlocal dielectric permittivity versus substrate temperature shown for the three values of wave vector $q$. The dead layer  dielectric constant is $\varepsilon_{\mathrm{DL}}=50$. For the large values of wave vector (green curve) the effective dielectric screening is very weak and almost independent on temperature. (b), (d) Relative renormalization of signle-layer MoS$_2$ bandgap as a function of dead layer permittivity for (b) different temperatures and (d) different thicknesses of dead layer. Panel (d) is plotted for room temperature. The strong renormalization can be attributed to weak screening of Coulomb interaction at the limit of large wave vectors.
    (c) Bandgap dependence on the substrate temperature. For larger values of dead layer permittivity the renormalization is smaller due to the reduced strength of Coulomb interaction. In all the panels we set the single-layer MoS$_2$ thickness $d=0.31$ nm, and dielectric permittivity $\varepsilon_{\rm TMD}= 6.4$ \cite{Laturia2018}. In panels (a), (b), (c) the dead layer thickness is $h=0.4$ nm. }
    \label{fig:bandgap}
\end{figure}

The parametric dependence of Coulomb interaction on the substrate temperature is reflected in the temperature-induced modulation of STO's dielectric constant given in Eq.~(\ref{eq:sto}). The heat impact on the effective dielectric function is shown in Fig. \ref{fig:bandgap}(a).
For small values of momentum, $q$, there is a well pronounced dependence of effective permittivity on temperature, mostly determining the excitonic structure. On the other hand, for larger values of $q$, the temperature dependence is negligible, and the value of permittivity is small, meaning a reduced screening of the Coulomb interaction. As we show later, this circumstance leads to strong bandgap renormalization due to Coulomb exchange between particles.
It should be also mentioned, that  the presence of the dead layer significantly suppresses the impact of STO giant permittivity.

\begin{figure}[t]
    \centering
    \includegraphics[width=1\linewidth]{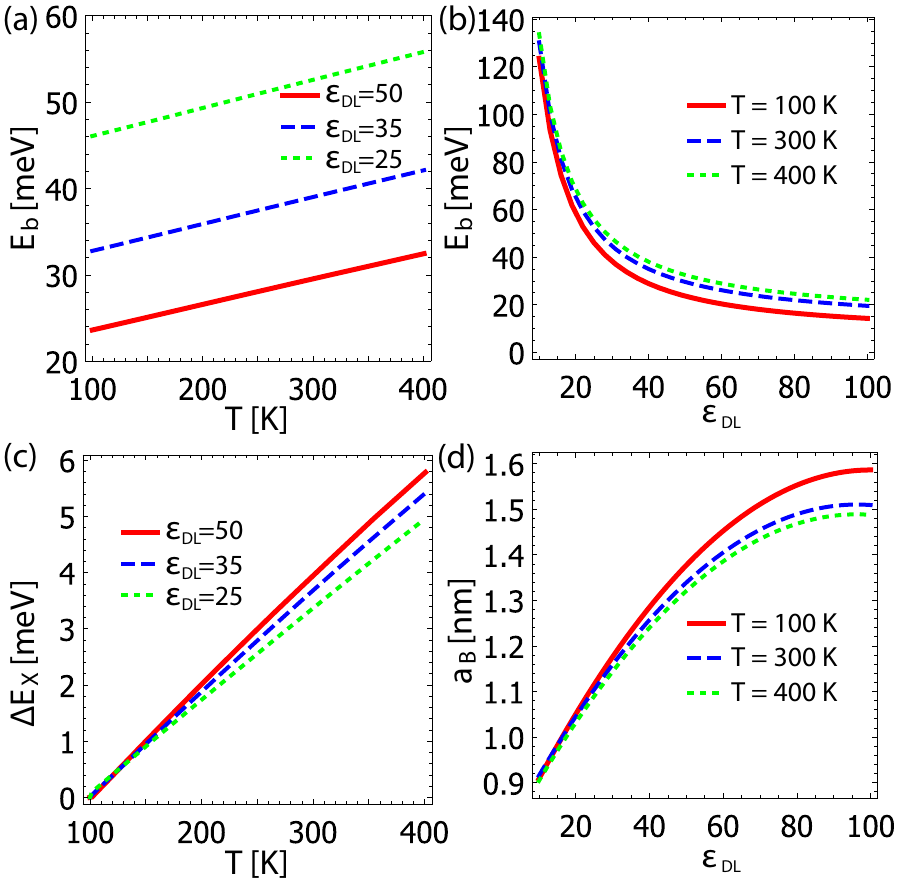}
    \caption{(a) Exciton binding energy as a function of substrate temperature for different values of dead layer permittivity. The modulation of Coulomb interaction strength results in variation of binding energy, while the presence of a dead layer makes the impact of substrate less pronounced. (b) Exciton binding energy and (d) Bohr radius as a function of dead layer permittivity at different substrate temperatures. (c) The shift of exciton resonant energy as a function of substrate temperature. Despite the similar dependence of  bandgap and exciton binding energy on substrate temperature, there is a significant shift of exciton resonance position. Here the dead layer thickness is $h=0.4$ nm, and  the parameters of single-layer MoS$_2$ are the same as in Fig. \ref{fig:bandgap}. }
    \label{fig:binding}
\end{figure}
\section{Results and discussion} \label{sec:result_discussion}
In this section we report and discuss our numerical results on the impact of substrate temperature on excitons in single-layer TMD.
\subsection{Bandgap renormalization}
The modulation of electron-electron interaction strength due to the variation of screening rate can lead to the renormalization of single-particle bandgap in TMD layer. In order to evaluate the strength of bandgap renormalization, we calculate the mean-field self-energy correction to the energy spectrum in TMD.
In the low energy limit, the electron Hamiltonian of a single atomic layer takes the following two-band form around the corner of hexagonal Brillouin zone \cite{Rostami13a,Liu13,Andor15}
\begin{equation}\label{H_TMD}
  \hat{\cal H}_{\rm TMD} ({\bm k})= d_I({\bm k}) \hat I + {\bm d}(\bm k) \cdot \hat {\bm \sigma}
\end{equation}
where $\hat{I}$ and $\hat {\bm \sigma}=(\hat{\sigma}_x$, $\hat{\sigma}_y$, $\hat{\sigma}_z)$ are the identity matrix and the components of Pauli matrix, respectively. The components of the Hamiltonian \eqref{H_TMD} read
\begin{align}\label{Delta}
&d_I (\bm k)=-\mu+\lambda_I \tau_z s_z+A (a_0k)^2~,
\nonumber\\
&{\bm d}({\bm k})= \left(a_0 t_0 \tau_z k_x,a_0 t_0 k_y, \Delta+\lambda \tau_z s_z+ B (a_0k)^2\right)
\end{align}
where $\mu=0$ is the chemical potential. Here, $\tau_z=\pm 1$, $s_z=\pm 1$ stand for valley and spin degrees of freedom, respectively. Note that $a_0=a/\sqrt{3}$, where $a=3.16 \AA$ is the lattice constant of MoS$_2$. The numerical values of this $k \cdot p$ Hamiltonian parameters
are taken as $t_0 \approx 2.34$ eV, $B\approx-1.135$ eV, $A\approx 121$ meV, $\lambda\approx-40$ meV, $\lambda_I\approx34.5$ meV and the bare bandgap is set as $E_{\rm g}=2\Delta \approx 2.6$ eV based on experimental \cite{Ulstrup2016} and theoretical \cite{Qiu2013} evidences.
The self-consistent mean-field self-energy correction is given by
\begin{equation}\label{self0}
  \hat{\Sigma}(\bm k)=- \frac{1}{\beta} \sum_{\bm q} \sum_{i\omega_n} V({\bm q-\bm k}) \hat{G}(\bm q,i\omega_n)~.
\end{equation}
Notice that $\beta=1/k_{\rm B}T$ with $k_{\rm B}$ as the Boltzmann constant, $i\omega_n$ is the fermionic Matsubara frequency, $V({\bm q})$ is the electron-electron interaction with a static screening effect and the Green's function follows
\begin{equation}\label{f_Green}
  \hat{G}(\bm q,i\omega_n) =\left[ i\omega_n -\hat{\cal H}_{\mathrm{TMD}}(\bm k) -\hat{\Sigma} (\bm k) \right]^{-1}~.
\end{equation}
We seek the self-energy function in the following generic two-band form
\begin{equation}\label{self1}
\hat\Sigma({\bm k}) = \Sigma_I({\bm k}) \hat I + {\bm \Sigma}({\bm k})\cdot \hat{\bm \sigma}~.
\end{equation}
Calculating the Green's function, performing Matsubara summation over the fermionic frequency, and considering the effective dielectric function given in Eq.~(\ref{eq:epsilon_eff}), one obtains the following self-consistent equations of the self-energy:
\begin{align}\label{self_system}
{\bm\Sigma} ({\bm k},T) \approx  \frac{1}{2} \sum_{\bm q}
\frac{v^{ee}_{|\bm q-\bm k|}}{\varepsilon_{\rm eff}(|{\bm q}-{\bm k}|,T)}
\frac{{\bm d}({\bm q})+ {\bm\Sigma} ({\bm q},T) }{|{\bm d}({\bm q})+ {\bm\Sigma} ({\bm q},T)|}
\end{align}
in which we explicitly note the $T$-dependence of the self-energy and $v^{ee}_q=2\pi e^2/q$ stands for the bare Coulomb interaction between two electrons. The approximation in Eq.~\eqref{self_system} corresponds to using the approximate values of Fermi-Dirac distribution functions as $f_c\approx 0$ and $f_v\approx 1$ for the conduction and valence band, respectively. This assumption is justified owing the large value of the band gap comparing to $k_{\rm B}T$ at room temperature.
Note that the $\Sigma_I$ term is decoupled from $\bm \Sigma$ when we use the approximate Fermi-Dirac function. In fact, it is shown that $\Sigma_I$ results in an irrelevant rigid energy shift~\cite{Borghi2009,Rostami15b} and therefore can be safely dropped. Moreover, we recall that $\sum_{\bm q} \equiv \int^{q_c}_0 \int^{2\pi}_0 q dq d\phi/(2\pi)^2$. We introduce an upper momentum cutoff $q_c$ chosen to preserve the total available phase space. Namely, considering the valley degeneracy $N_v=2$ we set $N_v \pi q^2_c=S_{\rm BZ}$, where $S_{\rm BZ} = 8\pi^2/(3\sqrt{3}a^2_0)$ is the area of the Brillouin zone. This implies $a_0 q_c =\sqrt{4\pi /(3\sqrt{3})} \approx 1.555$.
Eventually, the renormalized bandgap reads
\begin{equation}\label{E_b_renorm}
  E^\ast_{\rm g}(T) = E_{\rm g}+2\Sigma_z (0,T).
\end{equation}
The results of numerical simulations are shown in Fig. \ref{fig:bandgap}. One general observation is that the presence of Coulomb interactions essentially modifies bandgap.
Depending on the parameters of the dead layer, the renormalization rate, i.e. $(E^\ast_{\rm g}-E_{\rm g})/E_{\rm g}$, varies in 12-25$\%$ range [see Figs. \ref{fig:bandgap} (b), (d)]. In addition, there is a pronounced dependence on the substrate temperature, resulting in change of bandgap of about 15 meV for the temperature variation of 300 K.
Our numerical results based on the simple low-energy modeling are in a good agreement with previous sophisticated theoretical analysis and experimental measurements \cite{Qiu2013,Ugeda2014}.
\subsection{The modulation of binding energy}
The modulation of the substrate dielectric constant  $\varepsilon_{\rm STO}$ leads to modification of the spatial properties of excitons, which become coordinate-dependent. For a long wavelength temperature profile, we can consider that $\varepsilon_{\rm STO}$ changes smoothly in space. Therefore, any significant variation happens on the scale of tens of nanometers, while the Coulomb interaction inside the exciton occurs at sub-nm scale. This means that one can safely neglect the modulation of substrate permittivity as a function of exciton relative coordinate, {\it i.e.} ${\bm \nabla}_{\bm r} \varepsilon_{\rm STO}(T({\bm R}))  \approx 0$. The latter allows for the factorization of an exciton wave function in the form $\Psi({\bm r},{\bm R})=\chi({\bm r},{\bm R})\psi({\bm R})$, leading to the separation of exciton internal and CM dynamics:
\begin{equation}
    \label{eq:ScrhoInt}
    \left[-\frac{\hbar^2 \nabla^2_{\bm r}}{2\mu}   +V({\bm r},T({\bm R}))\right] \chi({\bm r},{\bm R})= -E_b({\bm R})  \chi({\bm r},{\bm R}),
\end{equation}
and
\begin{equation}
    \label{eq:SchroCM}
 \left[ -\frac{\hbar^2 \nabla^2_{\bm R}}{2M} +E_g({\bm R}) -E_b({\bm R})    \right] \psi ({\bm R})= E_{\rm X}({\bm R}) \psi ({\bm R}),
\end{equation}
where $E_b({\bm R})>0$ is the exciton binding energy. As it follows from Eq. (\ref{eq:SchroCM}), $E_b({\bm R})$ plays a role of the effective potential energy for the exciton CM wavefunction.

We note that Eq.~(\ref{eq:ScrhoInt}) with the potential given by Eq.~(\ref{eq:CPot}) is not exactly solvable. To characterize the exciton internal state, we employ a variational approach \cite{Shahnazaryan2017}, using as a trial function the conventional 2D exciton wavefunction \cite{Yang_pra_1991} in momentum space:
\begin{align}
\chi(q,{\bm R})= \frac{2\sqrt{2\pi} \lambda({\bm R})}{ (1+q^2 \lambda({\bm R})^2 )^{\frac{3}{2}}}~.
\end{align}
Here, $\lambda({\bm R})$ is a variational parameter and its critical value which minimizes the binding energy is the Bohr radius, {\it i.e.} $a_{\rm B} \equiv \lambda|_{\min[E_b]}$. The results of the binding energy calculation as a function of substrate temperature (dielectric permittivity) are presented in Fig. \ref{fig:binding}(a).
One immediate consequence is that the absolute scale of binding energy is much lower than in conventional hexagonal boron nitride (hBN) based setups, where it lies in 500 meV range \cite{Hill2015}. The latter arises from the strong screening of interaction, stemming from the enhanced impact of substrate permittivity.
It is notable, that while the dead permittivity sufficiently determines the absolute value of exciton binding energy, it almost does not modify temperature dependence. The latter demonstrates of about 10 meV change for binding energy in the corresponding temperature variation range for different values of dead layer permittivity. The values of exciton Bohr radius vary little with temperature, and can be approximated as $a_{\rm B} \approx 1.35; \quad 1.25; \quad 1.15$ nm for dead layer permittivity values $\varepsilon_{\mathrm{DL}} =50; \quad 35; \quad 25$, respectively. In contrast to binding energy, these values are quite close to exciton Bohr radius in TMD setups on conventional substrates (about 1 nm). The violation of the conventional ratio between exciton energy and Bohr radius $E_b\propto 1/a_{\rm B}$ is explained by the evidence that the Coulomb interaction generally in TMD based setups and particularly in the considered structure essentially differs from that in dielectrically homogeneous bulk media.

The dependence of exciton binding energy on dead layer permittivity is further revealed in Fig. \ref{fig:binding} (b), demonstrating almost an order of magnitude change for dead layer permittivity variation in the range of 10 to 100. The corresponding variation of Bohr radius is depicted in Fig. \ref{fig:binding} (d).
We also note that the binding energy can be easily fitted using a linear function:
\begin{equation}
    \label{eq:Binding_fit}
    E_b (T)\approx E_0 +\xi T,
\end{equation}
where the values of $E_0$ and $\xi$ depend on width and permittivity of the dead layer.
\begin{figure}
    \centering
    \includegraphics[width=1.0\linewidth]{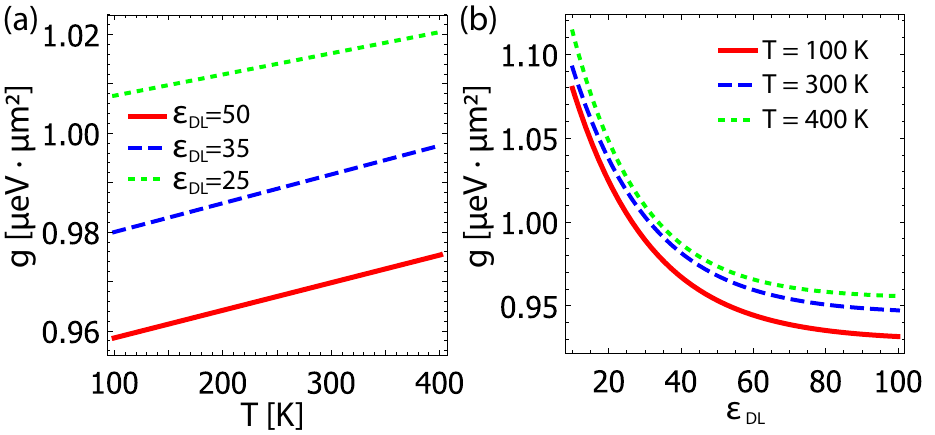}
    \caption{ (a) Exciton-exciton interaction strength dependence on substrate temperature.  (b) Exciton-exciton interaction as a function of dead layer permittivity. The reduction of Coulomb interaction strength with the increase of dielectric screening is partially compensated by the increase of exciton Bohr radius, resulting in  minor variation of interaction strength. }
    \label{fig:inter}
\end{figure}
%
\subsection{Exciton resonance energy}
The temperature dependence of exciton resonant energy shift relative to reference position at  $T=100$ K is shown in Fig. \ref{fig:binding} (c). The figure illustrates that while the variation of Coulomb interaction strength modifies both bandgap and exciton binding energy in the same direction, there is no exact cancellation between these effects, leading to a significant shift in the exciton resonance energy. The latter opens a way for exciton routing given that a temperature gradient throughout the sample is created. We also note that the obtained result is in qualitative agreement with the recent experimental observations of the exciton resonant energy response on the modulation of dielectric permittivity \cite{Lin2014,Raja2017,Hsu2019}.

\subsection{Exciton nonlinearity}
We proceed with the calculation of Coulomb interaction between excitons. Strictly speaking, the exact value of inter-exciton interactions (or the nonlinearity) depends on the particular shape of spatial dependence of binding energy, defining the form of exciton CM wave function. However, given by the smooth spatial variation of binding energy, one can assume the exciton-exciton interaction to be defined solely by the wave function describing exciton internal dynamics.
Within the scattering theory formalism, the inter-exciton interaction can be presented as a scattering event between two excitons of equal initial momenta, accompanied by momentum transfer from one exciton to another \cite{Ciuti1998}. The maximum of interaction appears at zero exchange momenta,  and in the explicit form reads  \cite{Glazov2009}:
\begin{align}
    g&=\frac{2}{A} \int V(q,T({\bm R})) \left[\chi \left({\bm k}+\frac{{\bm q}}{2} , {\bm R} \right)\right]^2 \chi \left({\bm k}-\frac{{\bm q}}{2} , {\bm R} \right) \notag \\
    &\left[\chi\left({\bm k}+\frac{{\bm q}}{2} , {\bm R} \right) -\chi\left({\bm k}-\frac{{\bm q}}{2} , {\bm R} \right) \right]
    \frac{d^2 {\bm q}}{(2\pi)^2} \frac{d^2 {\bm k}}{(2\pi)^2},
    \label{eq:inter}
\end{align}
where $A$ is the area of the sample. Here we neglect the contribution of direct dipole-dipole interaction, as for TMD monolayers it was shown to be negligibly small \cite{Shahnazaryan2017}.

The results of the calculation as a function of temperature are presented in Fig. \ref{fig:inter}(a). Similar to the binding energy, the interaction rate grows monotonously by increasing the temperature owing to the reduction of screening effect. The presence of dead layer makes the temperature dependence of interaction less pronounced, yet resulting in a few percent changes at 300 $K$ range. However, it should be mentioned that in striking difference with the binding energy, the absolute value of exciton-exciton interaction strength is not dramatically modified due to the giant screening by STO substrate. Particularly, considering a single-layer  MoS$_2$ deposited on hexagonal boron nitride (hBN) with permittivity $\epsilon_{\mathrm{hBN}}=5$, an exciton has the properties of $E_b^{\mathrm{hBN}}=344$ meV, $a_B^{\mathrm{hBN}}=1$ nm, $g^{\mathrm{hBN}}=1.32$ $\mu$eV$\cdot\mu$m$^2$. Such a behavior of the exciton-exciton interaction constant is in excellent agreement with an earlier performed investigation for WS$_2$ single-layer  \cite{Shahnazaryan2017}, where it was shown a reduction of interaction strength of 30$\%$ in the limit of vanishing screening length, corresponding to giant substrate screening of Coulomb interaction.
The small modulation of interaction energy stems from the fact that for relatively larger values of momentum, relevant for the integration in Eq.~(\ref{eq:inter}), the temperature dependence of Coulomb is very weak [see Fig. \ref{fig:bandgap}(a)]. The panel (b) in Fig.~\ref{fig:inter} illustrates the dependence of exciton-exciton interaction on the dead layer permitivitty.
The moderate change of interaction strength here can be attributed to the counterplay between the reduction of Coulomb interaction for the case of larger screening and the enhancement of exciton Bohr radius.
\subsection{Exciton routing}
\begin{figure}
    \centering
    \includegraphics[width=1\linewidth]{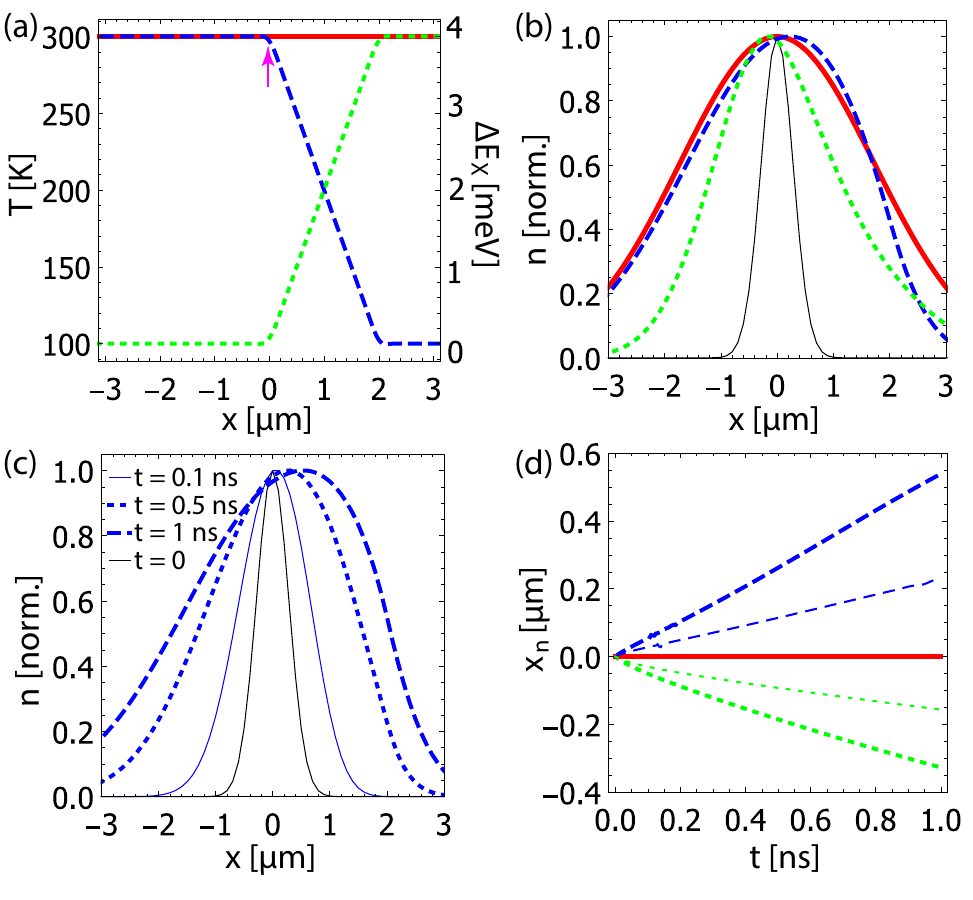}
    \caption{
    (a) Real space dependence of the substrate temperature. The right axes stands for corresponding change of exciton resonant energy.
    The magenta arrow indicates the region of localized resonant excitation of exciton cloud. (b) Normalized exciton density distribution at $t=1$ns after the excitation. The different colors correspond to the temperature profile in panel (a). Black curve shows the initial excitation spot.
    (c) Normalized exciton density distribution at different times for varying spatial profile of temperature, corresponding to the blue curve in panel (a).
    (d) The temporal dependence of maximum of exciton cloud. Thick [thin] lines correspond to the presence [absence] of Seebeck effect. The colors correspond to that in panel (a). The reduction [enhancement] of exciton energy creates attractive [repulsive] force, blue [green] curve. The Seebeck effect essentially enhances the routing efficiency.
    }
    \label{fig:routing}
\end{figure}
%
%
%
In the following subsection we consider the dynamics of exciton gas in a spatially inhomogeneous medium. For this purpose by means of inhomogeneous heating we create a spatially varying profile of substrate temperature. The latter results in corresponding spatial modulation of the exciton resonance energy.
We note here that given by the low exciton mobility and mean free path, Eq. (\ref{eq:SchroCM}) fails to describe the collective motion of the exciton cloud. Instead, the correct description of exciton propagation implies a diffusive treatment \cite{Kulig2018}, where the gradient of exciton resonance energy
leads to the appearance of a drifting force. In addition, the variation of temperature results in the modulation of diffusion coefficient as well. The diffusion coefficient is linked with the exciton mobility $\mu_{\rm X} $ via Einstein relation, $D(x)=\mu_{\rm X}  k_{\rm B} T(x)$. Here, exciton mobility is defined as $\mu_{\rm X} =\tau_c /M$, where $\tau_c$ is the average time interval between exciton collisions. For the monolayer MoS$_2$ it is found to be $\tau_c=260$ fs \cite{Cadiz2018}.
Thus, the exciton transport in the considered structure can be described by the following drift-diffusion equation:
\begin{align}
    \label{eq:diff_drift}
\left(\frac{\partial}{\partial t} +\frac{1}{\tau_{\rm X}} \right) n(x,t)= \mu_{\rm X} \left( \nabla^2 E_{\rm X} + k_{\rm B} \nabla^2 T_{\rm X} \right)  n(x,t)  \notag \\
+\mu_{\rm X} \left( \nabla E_{\rm X} + k_{\rm B} \nabla T_{\rm X} \right) \nabla n(x,t) +\nabla [D(x)\nabla n(x,t)],
\end{align}
where $\tau_{\rm X}$ is the exciton lifetime. The first terms in brackets on the right hand side of Eq. (\ref{eq:diff_drift}) stems from the spatial gradient of exciton resonant energy. The second terms correspond to the contribution of Seebeck effect \cite{Glazov2019, Perea-Causin2019}. The latter drives excitons from hot to cold areas of the structure, thus working in line with the considered routing mechanism. Here the impact of Seebeck mechanism is accounted for in the assumption that the temperature of exciton gas coincides with that of the structure, corresponding to the limit of fast exciton thermalization. It should be mentioned, that the dynamics of exciton propagation can be influenced by the impact of exciton-phonon coupling \cite{Glazov2019}, consideration of which however is beyond the scope of the current investigation.

The low-temperature radiative lifetime value for single-layer TMD deposited on an hBN substrate is about 1 ps \cite{ChernikovReview}. In contrast, at room temperature it  can reach up to $\tau\approx 1$ ns \cite{Shi2013,Palumno2013,Marie2016}. Yet, given  the reduced binding energy in the considered structure, one should account for the impact of non-radiative decay channels as well. For numerical simulations, we use the value of $\tau=100$ ps.

We consider three scenarios of spatial trend of substrate temperature and the corresponding exciton resonance energy relative to the initial excitation spot. These scenarios are depicted in Fig. \ref{fig:routing} (a), with an arrow indicating the position of the initial excitation of exciton cloud. For the sake of simplicity we choose linear dependence of the temperature spatial profile, meaning that the first bracket on the right hand side of Eq. (\ref{eq:diff_drift}) vanishes.
We start with the initial exciton distribution being $n(x,0)=n_0 \exp(-x^2/x_0^2)$, where $x_0=400$~nm, in accordance with relevant experimental conditions \cite{Kulig2018}. Figure \ref{fig:routing} (b) demonstrates the normalized exciton density spatial distribution at 1ns after initial excitation for different scenarios. In the presence of temperature gradient the combined action of exciton resonant energy difference and Seebeck effect pushes the maximum of the exciton cloud towards a colder area of the sample. \ref{fig:routing} (c) illustrates the snapshots of spatial profiles of exciton density at various moments after the initial excitation. The spatial gradient of temperature and exciton resonant energy here correspond to the blue curve in Fig. \ref{fig:routing} (a).

Finally, we study the contributions of resonant energy gradient and Seebeck effect due to the temperature gradient to the routing efficiency.
Fig. \ref{fig:routing} (d) illustrates the temporal evolution of the maximum of exciton distribution position accounting for Seebeck effect (thick lines) and in the absence of it (thin lines).
In principle a scenario when the Seebeck effect is absent can be realized when the gradient of exciton resonant energy is created by means of inhomogeneous strain of substrate rather than temperature gradient. However this task is beyond the scope of current paper and is left for future investigations.
As it follows from Fig. \ref{fig:routing} (d), the presence of Seebeck effect essentially enhances the efficiency of routing.

\section{conclusion}\label{sec:conclusion}
In conclusion, we studied the impact of a paraelectric substrate (using strontium titanate, SrTiO$_3$ as an example) on the excitonic properties of a TMD ({\it e.g.} MoS$_2$) monolayer. It was found that the giant screening of Coulomb interaction leads to sufficient bandgap renormalization and quenching of the exciton binding energy. The pronounced dependence of SrTiO$_3$ permittivity on temperature allows for rigid manipulation of exciton resonance energy in a single layer. By applying inhomogeneous heating, it is possible to create a prominent gradient of substrate temperature on the scale of tens of nanometers, directly mapped to the exciton resonant energy. The latter opens a possibility to create controllable exciton currents, having significant potential for exciton transport-based applications. Finally, we note that the proposed effect of exciton routing can be essentially enhanced for the case of strong light-matter coupling regime, leading to the formation of exciton-polaritons~\cite{Dufferwiel2017,Lundt2017}. The ultra-small effective mass of the latter, stemming from the photonic counterpart, would enhance the particle mobility by several orders, allowing to span the routing distance to tens and hundreds of $\mu$m.
\section*{Acknowledgements}
The authors are grateful to I. Chestnov, A. Balatsky and I. Shelykh for valuable discussions.
The work was supported by VILLUM FONDEN via the Center of
Excellence for Dirac Materials (Grant No. 11744).
H.R. acknowledges the support from the Swedish Research Council (VR 2018-04252).
V.S. acknowledges the support from Goszadanie No. 3.2614.2017/4.6 and Megagrant 14.Y26.31.0015 of the ministry of education and science of Russian Federation, and RFBR Grant No. 18-32-00873. The work was partly supported by the Government of the Russian Federation through the ITMO Fellowship and Professorship Program.

\end{document}